\documentclass[12pt]{iopart}
\pdfoutput=1
\expandafter\let\csname equation*\endcsname\relax
\expandafter\let\csname endequation*\endcsname\relax
\usepackage{mathtools} % which loads amsmath as well.
\usepackage[utf8x]{inputenc}
\usepackage{epsfig}
\usepackage{color}
\usepackage[toc,page]{appendix}
\usepackage{float}
\usepackage{graphicx}
\usepackage{anysize}
\usepackage{xfrac}
\usepackage{soul}
\soulregister\cite7
\soulregister\ref7
\soulregister\pageref7

\usepackage{soul}

\marginsize{3cm}{3cm}{2.5cm}{4.5cm}
\begin{document}

\title[About thermometers and temperature]{About thermometers and temperature}

\author{M. Baldovin}
\address{Dipartimento di Fisica, Sapienza Universit\`a di Roma, p.le A. Moro 2, 00185 Roma, Italy}
\ead{marco.baldovin@roma1.infn.it}
\author{A. Puglisi, A. Sarracino}
\address{CNR-ISC and Dipartimento di Fisica, Sapienza Universit\`a di Roma, p.le A. Moro 2, 00185 Roma, Italy}
\ead{andrea.puglisi@roma1.infn.it, alessandro.sarracino@roma1.infn.it}
\author{A. Vulpiani}
\address{Dipartimento di Fisica, Sapienza Universit\`a di Roma and CNR-ISC, p.le A. Moro 2, 00185 Roma, Italy}
\address{Centro Interdisciplinare B. Segre, Accademia dei Lincei}
\ead{angelo.vulpiani@roma1.infn.it}
\vspace{10pt}

\begin{abstract}
  We discuss a class of mechanical models of thermometers and their
  minimal requirements to determine the temperature for systems out of
  the common scope of thermometry. In particular we consider: 1)
  anharmonic chains with long time of thermalization, such as the
  Fermi-Pasta-Ulam (FPU) model; 2) systems with long-range interactions
  where the equivalence of ensembles does not always hold; 3) systems
  featuring absolute negative temperatures. We show that for all the
  three classes of systems a mechanical thermometer model can be
  designed: a temporal average of a suitable mechanical observable of
  the thermometer is sufficient to get an estimate of the system's
  temperature. Several interesting lessons are learnt from our
  numerical study: 1) the long thermalization times in FPU-like
  systems do not affect the thermometer, which is not coupled to
  normal modes but to a group of microscopic degrees of freedom; 2) a
  thermometer coupled to a long-range system measures its
  microcanonical temperature, even at values of the total energy where
  its canonical temperature would be very different; 3) a thermometer
  to read absolute negative temperatures must have a bounded total
  energy (as the system), otherwise it heavily perturbs the system
  changing the sign of its temperature. Our study shows that in order
  to work in a correct way also in ``non standard'' cases, the proper
  model of thermometer must have a special functional form, e.g. the
  kinetic part cannot be quadratic.
\end{abstract}

\section{Introduction}

An instrument able to measure a system's temperature must equilibrate
with that system within an acceptable time and without perturbing it
significantly. Among the system's properties which are usually
considered relevant for appropriate equilibration, one recognizes: 1)
a reasonable (not mathematical) notion of ergodicity, 2) the
possibility to take a small part of the system as a good
representative of the whole system, and 3) thermalization, in a suitable time,
among different parts of the system~\cite{huang,PSV17}.

As known, those requirements are satisfied by a widespread class of
systems in condensed matter physics, a fact which explains why
thermometers are common tools in everyday life and in scientific
applications~\cite{chang04}. Statistical physics, however, has indicated a series of
interesting systems where - for different reasons related to
particular forms of the interactions or constraints on kinetic and
potential energies - the fulfillment of the above simple requirements
may be complicated.

Chains of weakly anharmonic oscillators are a paradigm of slow - or
even not reached - equilibration: in this category the prototype
system is the Fermi-Pasta-Ulam (FPU) chain, which played a seminal
role in the history of chaos, numerical computation in
physics~\cite{FPU,galla2007,R2}, as well as in integrable
systems~\cite{R3}, and, in addition, is widely used also in non
equilibrium statistical mechanics, e.g. for the study of Fourier's
law~\cite{lepri2003}.  In this system, roughly we can say that
  for any finite $N$, if the energy per particle $E/N$ is small
  enough, one can observe the proper statistical features only after a
  typical time $t_c$, which diverges as $N \to \infty$ and $E/N \to
  0$, see~\cite{benettin2013}.  Notwithstanding this well-established
  behavior, we remark here that for the purpose of defining a
suitable thermometer we need the system to be {\em already} at
equilibrium: our only requirement, therefore, is that the thermometer
- a small perturbation which can be out-of-equilibrium at the instant
of contact - exchanges energy with the system exploring, within an
acceptable time, its statistical distribution.

A different complication originates from long-range
interactions~\cite{R2014}: such a kind of forces prevent the usual
argument which makes the interaction energy between a sub-system and
the rest of the system negligible. As a result, canonical ensemble is
not directly derived as the natural ensemble for small parts of a
microcanonical one, and - conversely - equivalence of ensembles is no
more guaranteed~\cite{barre2001}. A priori it is not obvious that
  the equilibration times are reasonably small, even in an optimistic
  scenario. The crucial question here is: what is the temperature
{\em read} by a thermometer coupled to such a long-range system, at
energies for which microcanonical and canonical ensembles predict
different results?

The third category of complication considered below is the one coming
from systems with bounded energy: such a property - which is verified
also in certain experiments~\cite{Braun2013} - implies a
microcanonical entropy which is non-monotonous in the energy, and
therefore a range of energies where the absolute temperature is
negative~\cite{cerino15,Frenkel2014}. While negative temperatures do
not entail any paradox for thermodynamics, their existence is strictly
related to the bounded range of achievable energies. This means that
coupling to such systems a normal thermometer (e.g. with kinetic
energy of the usual kind $p^2/2m$) results in a catastrophic
change of the state of the system which, necessarily, must transfer a
very large amount of energy to the thermometer: in other words a
normal thermometer, even very small, represents for those systems a
huge perturbation. Experimental realization of systems with negative
temperature is in fact conditioned on the ability of isolating the
system from the rest of the world. In addition to such a sensitivity
to external perturbations, in systems with negative absolute
temperature the equipartition of energy is almost always broken~\cite{cerino15}, an
evident further complication in designing a suitable thermometer.

In the present paper we propose a general model of ``mechanical''
thermometer which, with small adaptations, is able to measure
temperature in examples of all the three classes mentioned above.
Let us stress that the Hamiltonian of the proper thermometer must
  have some specific form, see Sec. 3.3.  We enforce molecular
dynamics simulations of the systems and the coupled thermometer,
showing for the latter a dynamical evolution toward equilibration at
the correct temperature. In all our numerical experiments a certain
amount of noise in the measurements of temperature will be evident: we
remark that the observables which we measure are estimators of the
system's temperature and for this reason are subject to statistical
uncertainties, which of course cannot be considered as fluctuations of
temperature~\cite{FVVPS11}. We recall that a similar proposal of a
mechanical thermometer was introduced in the - quite different -
framework of aging systems with the aim of measuring the
Fluctuation-Dissipation effective
temperature~\cite{cugli2,exartier00}. Another class of systems, again
in the out-of-equilibrium realm, is constituted by granular
fluids~\cite{puglio15}, where the problem of measuring temperature by
means of probing a small sub-system has been studied
in~\cite{PBL02,BLP04}.

The paper is organized in the following way. In Section 2 we discuss
the classical definition of microcanonical temperature, we recall a
precious formula allowing universal estimate of temperature through a
dynamical measurement, and we present our model of a thermometer. In
Section 3 we discuss numerical experiments where the thermometer is
coupled to an FPU chain, a generalized Hamiltonian mean-field model
(example of a system with long-range interactions), and finally with a
model of coupled rotators with bounded kinetic energy, providing an
example of system with negative temperatures. In Section 4 we draw our
conclusions.

\section{What temperature is and how to measure it}

Many different types of efficient thermometers have been conceived,
often involving rather sophisticate technologies~\cite{lee2007}. The
main aim of the present paper is to discuss the conceptual aspects
behind the problem of measuring temperature. For this reason, of
course, we do not claim to improve the practical and technical
aspects of thermometry.  A first conceptual prerequisite to propose a
suitable candidate for a thermometer is a clear understanding of the
concept of temperature.

\subsection{Definitions of temperature}

From the basic principles of the statistical mechanics we know how to
define the microcanonical temperature $T$ in a system with Hamiltonian $H({\bf Q},{\bf P})$:
$$
\beta={1 \over k_B T}= {\partial \over \partial E} \ln \omega(E),
$$
where 
$$
\omega(E)=\int \delta(E-H({\bf Q},{\bf P})) d{\bf Q} d{\bf P}, 
$$ 
with $k_B$ the Boltzmann constant.
Such a definition, although rather important, does not appear very
useful for practical purposes. Measuring or computing the density in
phase space is usually impossible. A possible alternative is to
invoke the equipartition formula
\begin{equation}
\left\langle X_m \frac{\partial H}{\partial X_n}\right\rangle  = \delta_{mn} k_B T,
\label{eq1}
\end{equation}
where $\langle\cdots\rangle$ indicates the average with respect to the
microcanonical measure. On the other hand - as discussed
in~\cite{cerino15} - such formula does not always hold in certain
kinds of systems, for instance those with negative temperatures.  The
possible failure of Eq.~(\ref{eq1}) is mainly a consequence of bounded
potential energy~\cite{cerino15}.

A precious help, for this task, is obtained following the approach
introduced by Rugh~\cite{rugh97,rugh98}. According to it, we can
determine $\beta$ through the time average of a function:
\begin{equation}
\beta={1 \over k_B T}= \left< \Phi ({\bf X}) \right> \,\, , \,\,  {\bf X}=({\bf Q},{\bf P}),
\label{rugh}
\end{equation}
where
$$
\Phi=\nabla \cdot \Bigg( {\nabla H \over || \nabla H ||^2} \Bigg).
$$
Rugh's approach is very elegant and rather relevant from a conceptual
point of view. In fact,

\begin{itemize}

 \item  it gives us a  definition of temperature for any kind of  Hamiltonian system;

 \item formula~\eqref{rugh} allows for computation of the temperature
   as a time average of an observable, e.g. from molecular dynamics
   simulations, and, at least in principle, in real experiments.

\end{itemize}
We notice that the use of Eq.~\eqref{rugh} does not give particular
advantages in systems with the usual form of the Hamiltonian,
e.g. those with a quadratic kinetic part and a potential contribution,
where equipartition formula is a valid and perhaps simpler
alternative, see e.g.~\cite{giardina98}. On the contrary, in systems with
peculiar phase space structure, such that equipartition is not
guaranteed, formula~\eqref{rugh} becomes very important.

Let us now discuss another {\em general-purpose} way to define the temperature.
In a large system whose Hamiltonian has the form:
$$
H=\sum_{n=1}^N g({\bf p}_n) + V({\bf q}_1, .... , {\bf q}_N),
$$
it is easy to find  the probability density function of
the momentum with a generalized Maxwell-Boltzmann 
formula:
\begin{equation}
\rho({\bf p}) = const.\, e^{-\beta g({\bf p})}.
\label{eq2}
\end{equation}
The previous result allows for a practical way to determine $\beta$
from a time average of a suitable function of ${\bf p}$.  The most
common case is the familiar case $g(p)=p^2/(2m)$: $\beta^{-1}=\left<
p^2/m \right>$. The average $\langle \rangle$ here will be obtained
through a double average in time and on the particles of the
thermometers (or - for comparison - of the system). Remarkably
Eq.~(\ref{eq2}) holds also for systems with negative
temperature~\cite{cerino15}. In the present paper we will use
Eq.~(\ref{eq2}) as the starting point to measure temperature. Our
idea, however, is that the observer can measure observables only in
the thermometer and not directly in the system. In the following we
discuss how a thermometer can be modelled and coupled to the system of
interest.

\subsection{A minimal model for a  thermometer}

Our aim, here, is to introduce a very general setting for modelling
the act of measuring temperature, from a dynamical point of view. In
particular we need a mechanical model for a thermometer which can be adapted also to
the three special cases considered in the rest of the paper. Our
definitions are quite natural and do not reserve particular surprises.

We consider:
\begin{itemize}

\item a system with Hamiltonian $H_S({\bf X})$ where ${\bf  X}\in R^{N_S}$ denotes the system's mechanical variables and in addition $N_S
  \gg 1$;

\item a system (the ``thermometer'') with Hamiltonian $H_T({\bf Y})$ where ${\bf Y}\in R^{N_T}$ denotes the thermometer's mechanical
  variables and $N_T\ll N_S$ ($N_T$ can be small, in principle).

\end{itemize}
The whole system therefore has the following Hamiltonian:
$$
H_S({\bf X})+H_T({\bf Y})+\epsilon H_I({\bf X}, {\bf Y}),
$$
where we assume that the system weakly interacts with the thermometer i.e.
$\epsilon \ll 1$. As possible interactions we consider
$$
H_I=\sum_{i,j}V_{i,j}(q_i -Q_j),
$$ 
where $\{ q_i \}$ ($i=1,2,.. , N_T$) denote the positions of the
particles of the thermometer and $\{ Q_j \}$ ($j=1,2, .., N_S$) the
ones of the system. The particular forms of internal interactions and
kinetic energies (for the system and for the thermometer) will be
varied in the three numerical experiments explained below.

\section{Numerical computations on different systems}

We have put in evidence three classes of non-trivial systems (choices
of $H_S$) where the possibility to measure temperature by coupling a
thermometer is a priori challenged by some apparent complication.

\begin{itemize}

\item FPU-like systems. In such a class of models the validity of a
  fundamental assumption for the usual statistical mechanics
  (e.g. ergodicity) is not completely clear. 

\item Systems with long-range interactions. For these systems the
  equivalence of ensembles is not always guaranteed, resulting in a possibility of
  ambiguity for the expected value of temperature. 

\item Systems with negative temperature. Isolation is crucial for the
  survival of these systems, therefore the contact with a (standard) thermometer,
  even very small, could be dramatic. We will see that suitable
  thermometers exist also for this class.

\end{itemize}

Our numerical simulations follow the usual Verlet algorithm, with
time step chosen to be $5\times 10^{-3}$, in order to keep relative
energy fluctuations $< 10^{-4}$. In most of our numerical experiments
we have applied the following protocol: 1) we have initialized the
system in a typical thermal (equilibrium) state; 2) we have verified
dynamically that the system is at equilibrium; 3) we have coupled the
thermometer and 4) we have read the temperature averaging one or more
observables (as discussed in Sec. 2.1), see captions of the figures
for details. We underline that our numerical simulations are fully
deterministic, i.e. without coupling with external reservoirs or
thermostats.

\subsection{FPU chain}

Consider the usual  FPU model, i.e. a chain of weakly non linear oscillators:
$$
 H_S=\sum_{i=1}^{N_S}\frac{P_i^2}{2 M}+\sum_{i=1}^{N_S+1}\frac{1}{2} (Q_i-Q_{i-1})^2+\sum_{i=1}^{N_S+1}\frac{\alpha}{3} (Q_i-Q_{i-1})^3+\sum_{i=1}^{N_S+1}\frac{\beta}{4} (Q_i-Q_{i-1})^4
$$ 
with $Q_0=Q_{N_S+1}=0$, $\alpha$ and $\beta$ positive constants, and
$$
H_T=\sum_{i=0}^{N_T}\frac{p_i^2}{2 m}+\sum_{i=1}^{N_T}(q_i-q_{i-1})^2 \,\, , \,\, H_I=\sum_{i=1}^{N_T} \frac{1}{2}(Q_i-q_i)^2.
$$ 
Many studies (see for instance~\cite{lepri2003,galla2007,R2} and
references therein) give a strong evidence of the following scenario:
starting with initial conditions very far from the equilibrium
(typically the energy concentrated only on few low frequency normal
modes), if the energy per particle $E/N$ is not large enough, the
system is not able to reach the thermal equilibrium.  More
  precisely for small values of $E/N$ one can reach thermal
  equilibrium only after a time very long which diverges as $N \to
  \infty$ and $E/N \to 0$~\cite{benettin2013}.

Following the previous results some authors considered the problem of
the temperature in the FPU model, studying the features of the system
interacting with a thermometer and a thermal bath.  At variance with
our model of thermometer, the authors of
Refs.~\cite{galgani1992,benettin2007} considered the cases where just
one particle of the system interacts with the thermometer and, in
addition, the initial conditions are typically very far from the
thermal equilibrium. In such an approach, clearly in the tradition of
studies on FPU, the main interest is for the possible presence of
(very) long metastable behaviors whose relaxation times increase as
$E_S/N_S$ decreases.

\begin{figure}[ht!]
\centering
\includegraphics[width=0.49\linewidth]{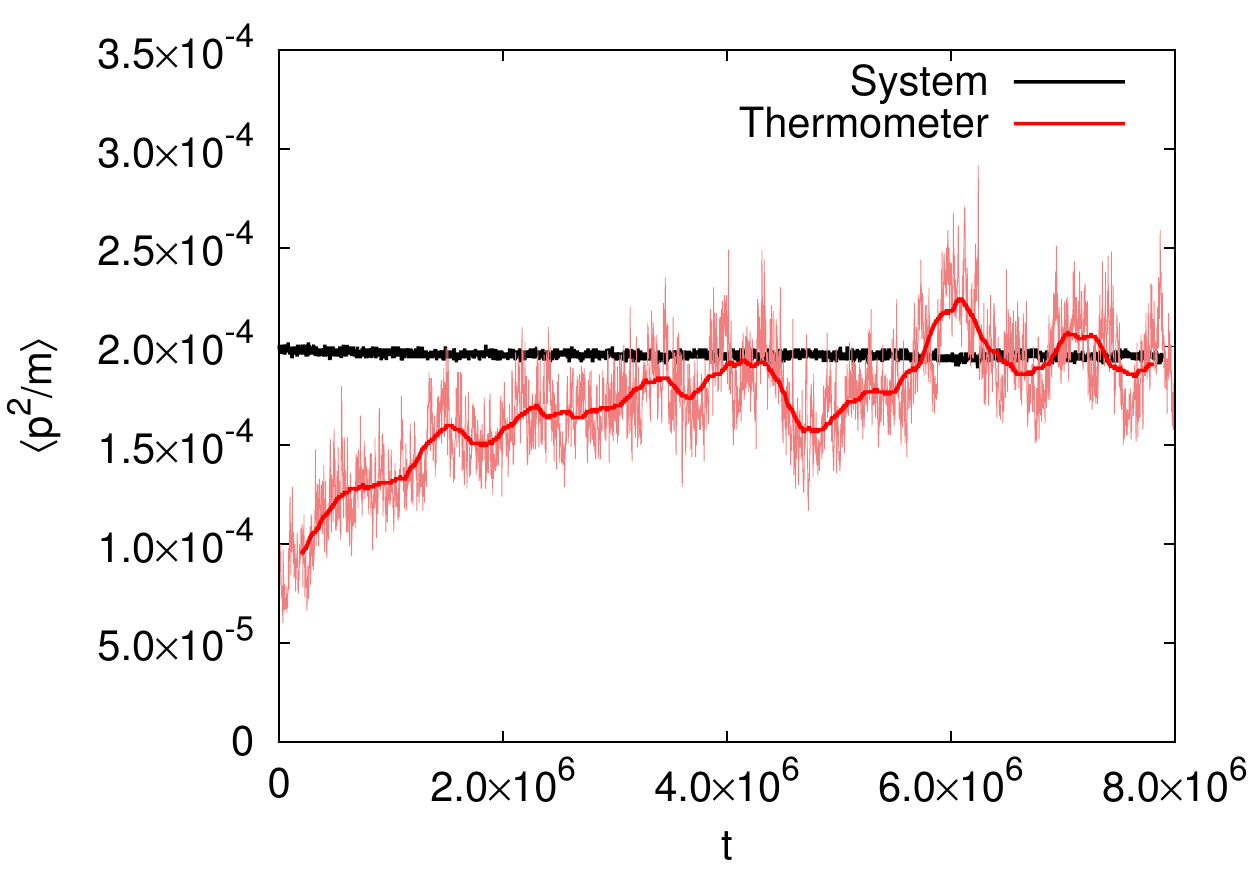}
\includegraphics[width=0.49\linewidth]{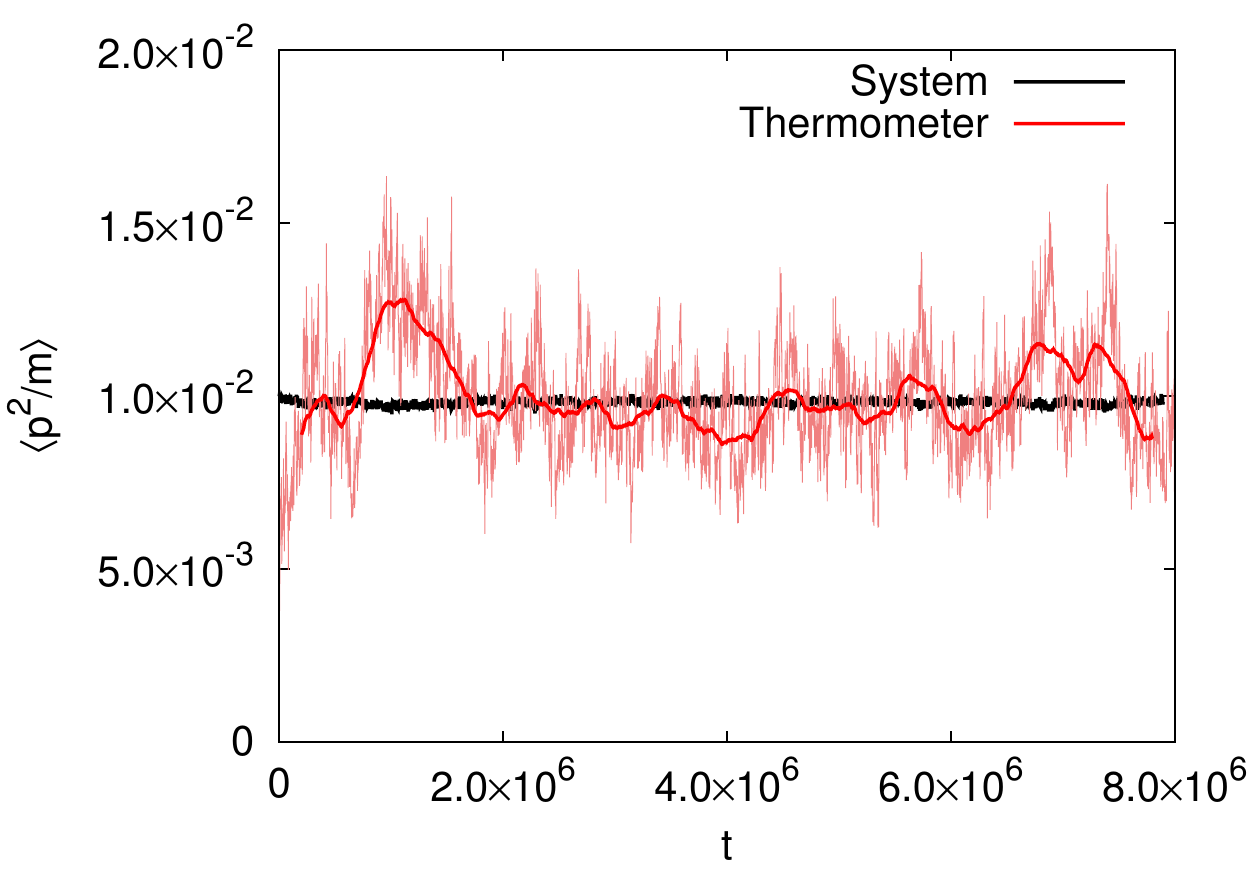}
\caption{Measuring temperature in FPU chains. Evolution of $\langle
  p^2/m\rangle$ for the system (black line) and the thermometer (red
  lines) after they have been connected, here $\langle p^2/m
    \rangle$ is averaged also over all the components of the systems
    (or the thermometer). We consider the cases
  $E_S/N_S=2\times10^{-4}$ (left) and $E_S/N_S=10^{-2}$ (right). For
  the thermometer, both ``instantaneous'' averages over $\Delta t=500$
  time units and moving averages on a time window $\tau=25000$ are
  shown. Parameters are $\alpha=1$, $\beta=2$, $M=1$, $N_S=1000$,
  $N_T=30$, $\epsilon=0.01$, and $m=0.2$.}
\label{fig:fpu_time}
\end{figure}

\begin{figure}[ht!]
\centering
\includegraphics[width=0.6\linewidth]{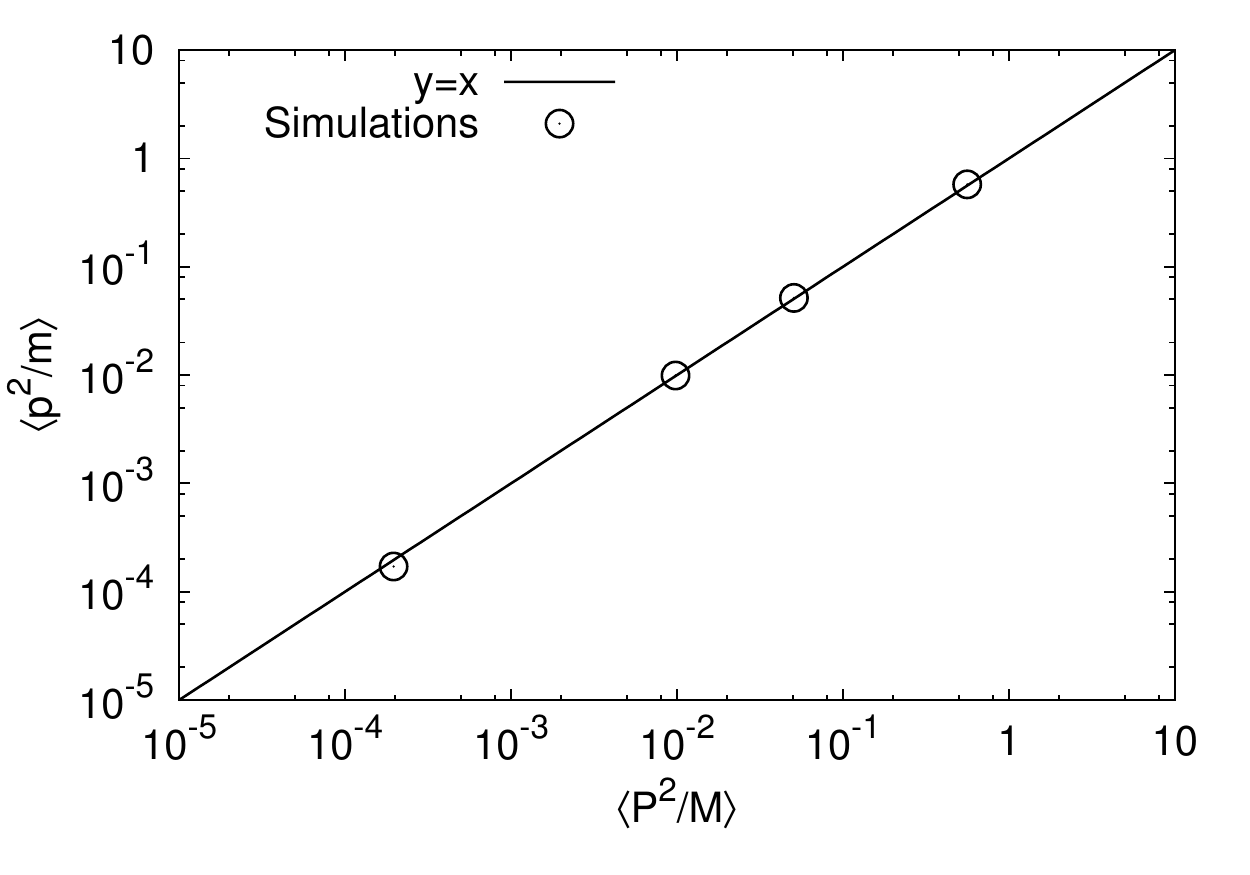}
\caption{Measuring temperature in the FPU chains. $\langle
  p^2/m\rangle$ as a function of $\langle P^2/M\rangle$. The
  parameters are as in Fig.~\ref{fig:fpu_time}. $E_S/N_S=2\times
  10^{-4},\,10^{-2},\,0.1\text{ and }1$, and $E_T/N_T=2\times 10^{-5}$
  at $t=0$.}
\label{fig:fpu}
\end{figure}

On the contrary here we are interested in a temperature
measurement in a standard situation, i.e. checking the ability of a thermometer to thermalize
with the system and bring information about its temperature. The
initial conditions for the system and the thermometer used in our
simulations are reported in the caption of Fig.~\ref{fig:fpu}.  The
results of our numerical simulations are reported in
Fig.~\ref{fig:fpu_time}, where we show the evolution with time of the
variance of the momentum distribution for the thermometer and in
Fig.~\ref{fig:fpu}, where we compare such a variance with the variance
of the chain's momentum distribution.  Let us note that in
Fig.~\ref{fig:fpu} the values of $E_S/N_S$ vary from below the
``critical'' threshold (where one has weak chaos and the failure of
equipartition~\cite{galla2007,R2}) to large energy per particle, for which the dynamics is
in good agreement with the statistical mechanics prediction.

In Fig.~\ref{fig:fpu_time} we see that the thermometer works
  rather well also at small energies (below the threshold), but the
  thermalization time is longer.  Basically we can say that the weak
ergodicity at low energy per particle does not produce particular
problems and the thermometer works in the proper way.  One can wonder
about the origin, in our simulations, of the absence of the
statistical anomalies observed in many numerical
studies~\cite{galla2007,R2}, e.g. the lack of equipartition with
initial condition very far from equilibrium.

Let us note that in the FPU system at low energy the normal modes are
almost decoupled. Nevertheless starting from a typical equilibrium
initial condition (as in our computation), for $N_S \gg 1$,
non-negligible fluctuations of the energy of a subregion can be
observed and therefore an exchange of energy between the system and
the thermometer can be realized: in particular, the thermometer is
able to work in the proper way.  We can say that our results are in
agreement with those of Khinchin on the pure role of the details of
the dynamics for the time average of global quantities in high
dimensional systems: thanks to this ``practical ergodicity'' the
thermometer is able to behave in the proper way also at low energy~\cite{Khinchin}.

\subsection{Long-range systems}

\begin{figure}[ht!]
\centering
\includegraphics[width=0.49\linewidth]{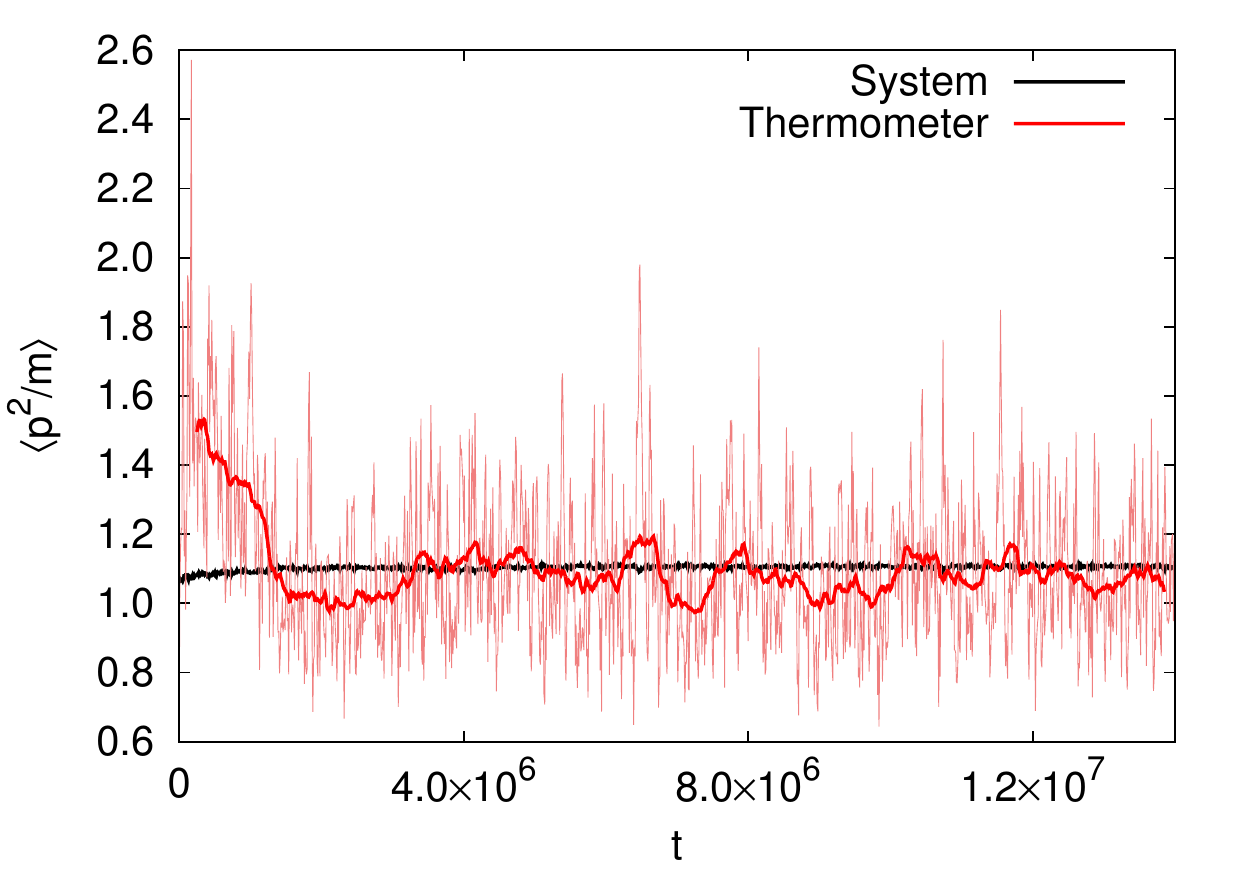}
\includegraphics[width=0.49\linewidth]{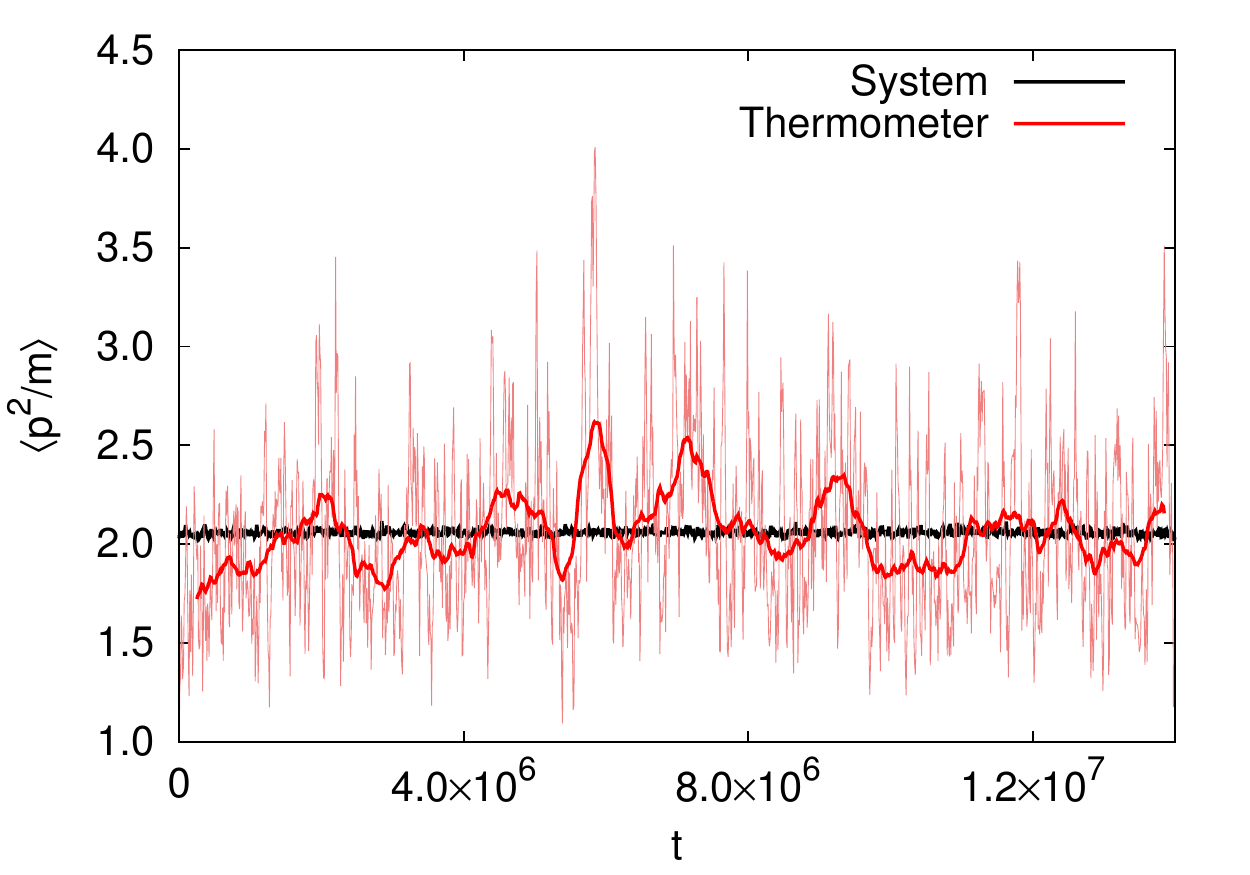}
\caption{Measuring temperature in
  long-range systems. Evolution of $\langle p^2/m\rangle$ for the
  system and the thermometer. Here we show the cases $E_S/N_S=1.2$ (left) and
  $E_S/N_S=3.2$ (right). As in Fig. \ref{fig:fpu_time}, both
  ``instantaneous'' ($\Delta t=1000$) and moving ($\tau=50000$) averages are
  shown. Parameters: $N_S=1000$, $N_T=30$, $\epsilon=0.1$, $J=1$,
  $K=10$, $m=0.2$, $\gamma=0.8$.}
\label{fig:lr_time}
\end{figure}

\begin{figure}[ht!]
\centering
\includegraphics[width=0.6\linewidth]{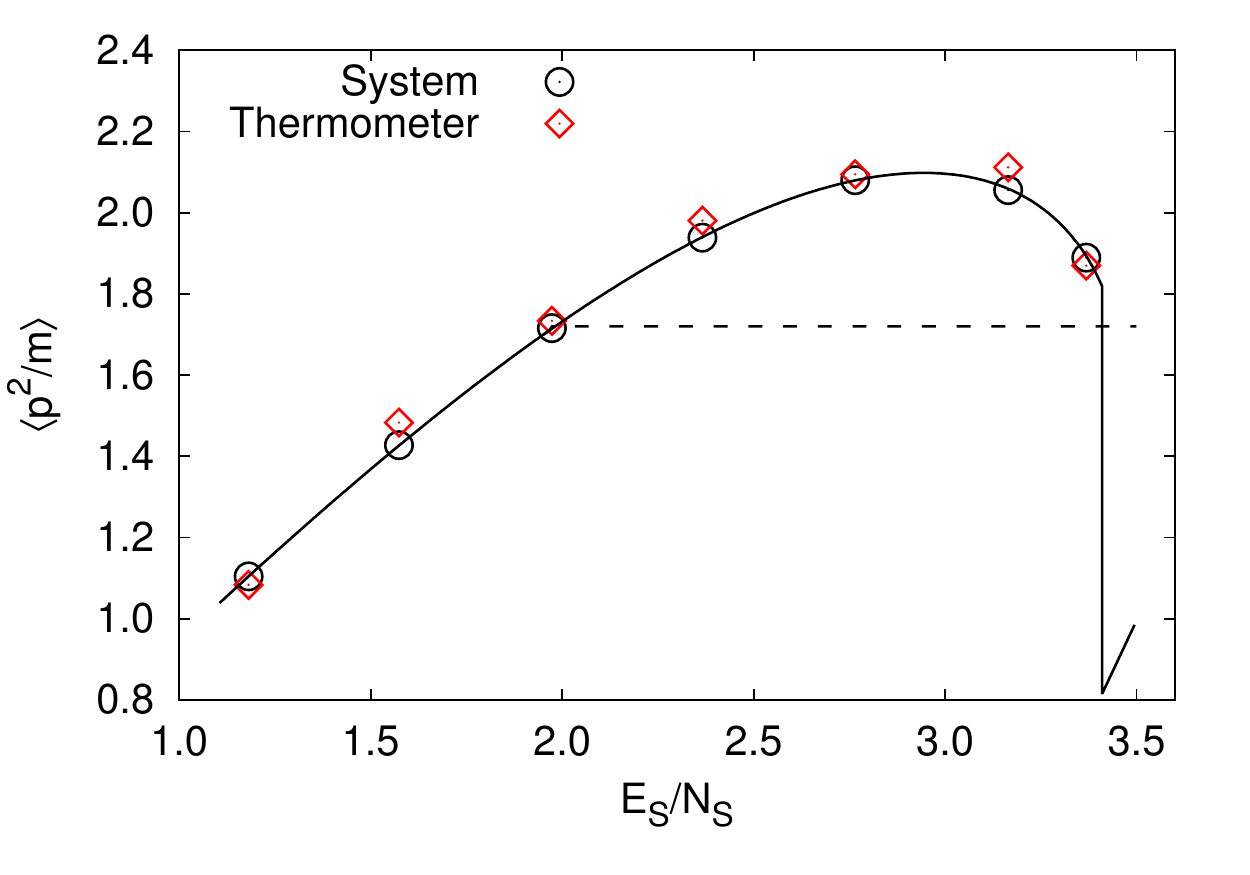}
\caption{Measuring temperature in long-range systems. $\langle p^2/m\rangle$
(red squares) and $\langle P^2/M\rangle$ (black circles) as functions of $E_S/N_S$.
The solid line shows the theoretical behaviour of a system in the microcanonical ensemble,
the dashed one is the result for the canonical ensemble obtained with the Maxwell construction. Same parameters
as in Fig. \ref{fig:lr_time}.}
\label{fig:lr}
\end{figure}
%fine mb

Usually statistical mechanics considers systems with short range
interactions.  Such a class of systems is physically rather important
(e.g. gases or liquids) and, in addition, it can be treated
mathematically showing, in a rigorous way, many relevant properties,
e.g. the existence of a thermodynamic limit and the equivalences of
the ensembles in the limit of very large systems.

On the other hand there are systems with long-range interactions,
which are very important for instance in fluid dynamics, laser and plasma physics,
and astrophysics~\cite{barre2001}.  In addition, in the presence of long-range
interactions one can have a failure of the equivalence of the
ensembles, which means that the results obtained in the microcanonical
and in the canonical ensembles can be different also in the
thermodynamic limit.

An interesting case of long-range interacting system in which
canonical and microcanonical ensembles are not equivalent is the
Generalized Hamiltonian Mean Field model (GHMF), introduced
in~\cite{ruffo1995}. The system is described by $N_S$ angular positions
$\theta_i \in [-\pi, \pi)$, $i=1,...,N_S$ and their conjugate momenta
  $P_i$. The Hamiltonian of the system can be written as:
\begin{equation}
\label{eq:lr}
H_S=\sum_{i=1}^{N_S}\frac{P_i^2}{2 } +N_S\frac{J}{2} (1-\mu^2)+ N_S\frac{K}{4}(1-\mu^4),
\end{equation}
where
\begin{equation}
 \mu=\sqrt{\mu_x^2+\mu_y^2}, \quad \mu_x=\sum_{i=1}^{N_S}\cos\theta_i, \quad \mu_y=\sum_{i=1}^{N_S}\sin\theta_i,
\end{equation}
and $K$ and $J$ are positive constants.
In our simulations one particle was always anchored in the origin by a
restoring force, in order to prevent rigorous conservation of the total momentum.
\\
Since the positions are angular variables, it is rather natural to introduce our thermometer with the same features, i.e.:
\begin{equation}
 \label{eq:term_ang}
 H_T=\sum_{i=0}^{N_T}\frac{p_i^2}{2 m}+\sum_{i=1}^{N_T}m\gamma^2\left[1-\cos(\phi_i-\phi_{i-1})\right],
\end{equation}
with $\gamma$ a positive constant.
Finally, we choose an interaction term such that the forces between the particles are periodic in $\phi_i-\theta_i$:
\begin{equation}
 H_I=\sum_{i=1}^{N_T} \left[1-\cos(\phi_i-\theta_i)\right].
\end{equation} 

As mentioned above, the system~(\ref{eq:lr}) in a range of values of $E_S/N_S$
shows a statistical behavior which is quite different from the usual
one. Namely, $E_S/N_S$ is a nonmonotonic function of the temperature
and the results obtained with the canonical and microcanonical
ensembles are different, even for $N_S\gg 1$. In
Fig.~\ref{fig:lr_time} we show the time evolution of the variance of
the single-particle distribution of momentum for the system and for
the thermometer. In Fig.~\ref{fig:lr} the time-averaged values of
those observables are shown for a wide range of energies,
demonstrating an overall agreement with the microcanonical
expectation, also for the temperature read by the thermometer.

\subsection{Systems with negative temperature}

In the present subsection we use the following  Hamiltonian 
\begin{equation}
\label{eq:tn}
H_S=\sum_{i=1}^{N_S}\left( 1- \cos P_i\right)+ \frac{J}{2 N_S} \sum_{i,j=1}^{N_S}\left(1-\cos(\theta_i-\theta_j)\right) + K \sum_{i=1}^{N_S}\left( 1- \cos (\theta_{i}-\theta_{i-1})\right)
\end{equation}
with $\theta_0=0$. We decided to use the above system for the following reasons:

\begin{itemize}

\item In the limit $J=0$, we have the model discussed in~\cite{cerino15}, showing absolute negative temperatures; it also corresponds to the model used to interpret the experiments in~\cite{Braun2013};

\item For $K=0$, one has a generalization of the GHMF
  model~\cite{ruffo1995}.

\end{itemize}

Let us open a short parenthesis on the above system.  For the FPU
system, the ``natural'' variables are the normal modes, which, even in
a statistical analysis, may show regular behavior and are able to
remember for a very long time the initial conditions: therefore the
approach to the equipartition can be, if any, very slow. However, even
if the normal modes are almost decoupled, when one looks at ``local''
variables $\{ q_n, p_n \}$, basically one recovers the correct
features predicted by statistical mechanics.  On the contrary, for the
chain of (generalized) rotators in Eq.~\eqref{eq:tn} with $J=0$ at
large energy the normal modes, i.e. the carriers of the energy, are
the local variables $\{ \theta_ n, P_n \}$ themselves, and therefore
exchange of energy among the subsystems is strongly
depressed~\cite{Livi1987}.  Therefore in order to avoid non ergodic
behavior, or more likely, very slow exchange of energy, a small non local
interaction contribution has been introduced (the term with $J$).
Such a term has the mere aim of avoiding long living metastable
situations. Let us also note that for this model, using a large
deviation approach, one can obtain an analytical prediction for the
behavior of $\beta$ as a function of $E/N$.  The results of our
computations are reported in the Appendix.

\begin{figure}[ht!]
\centering
\includegraphics[width=0.4\linewidth]{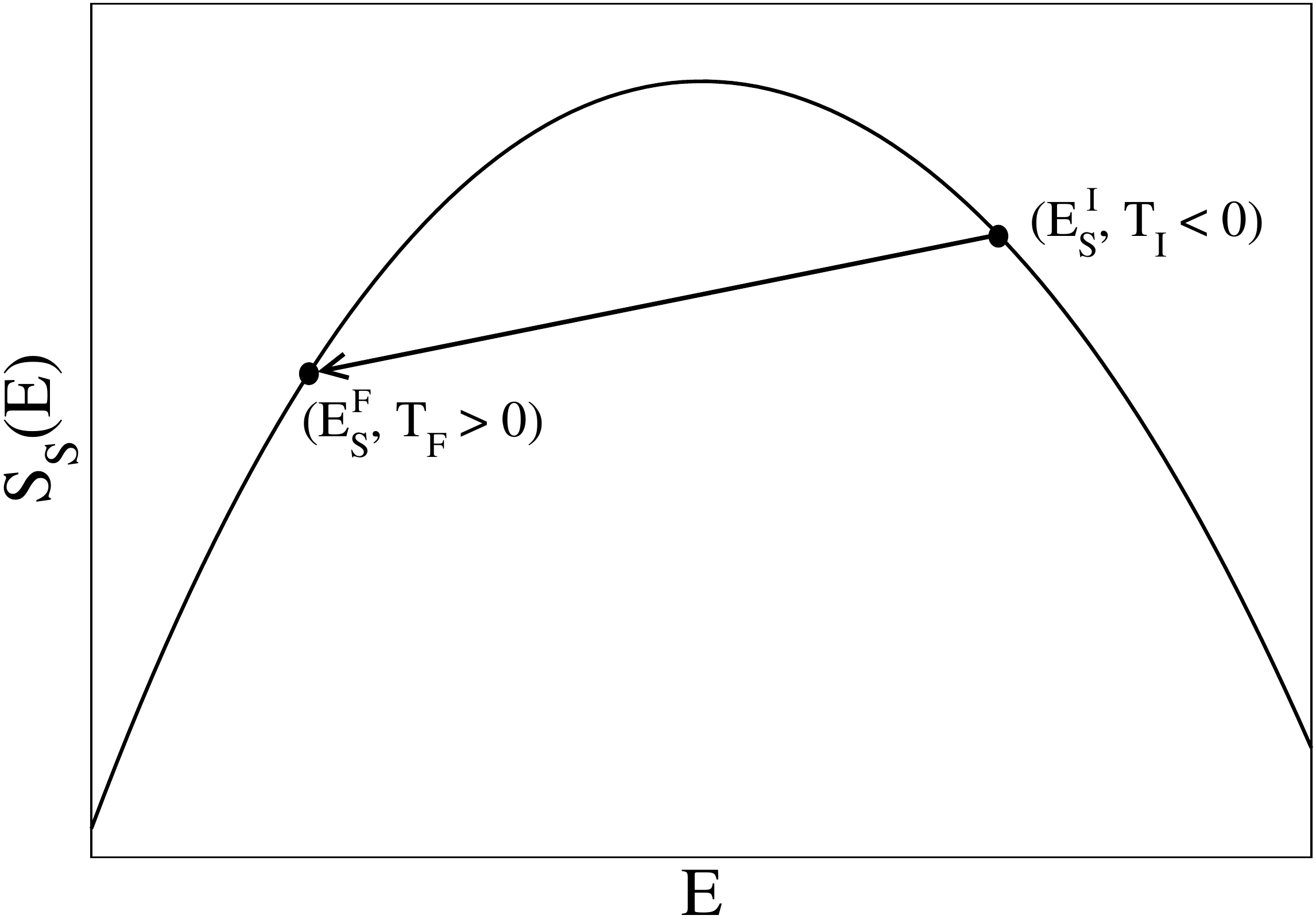}
\includegraphics[width=0.4\linewidth]{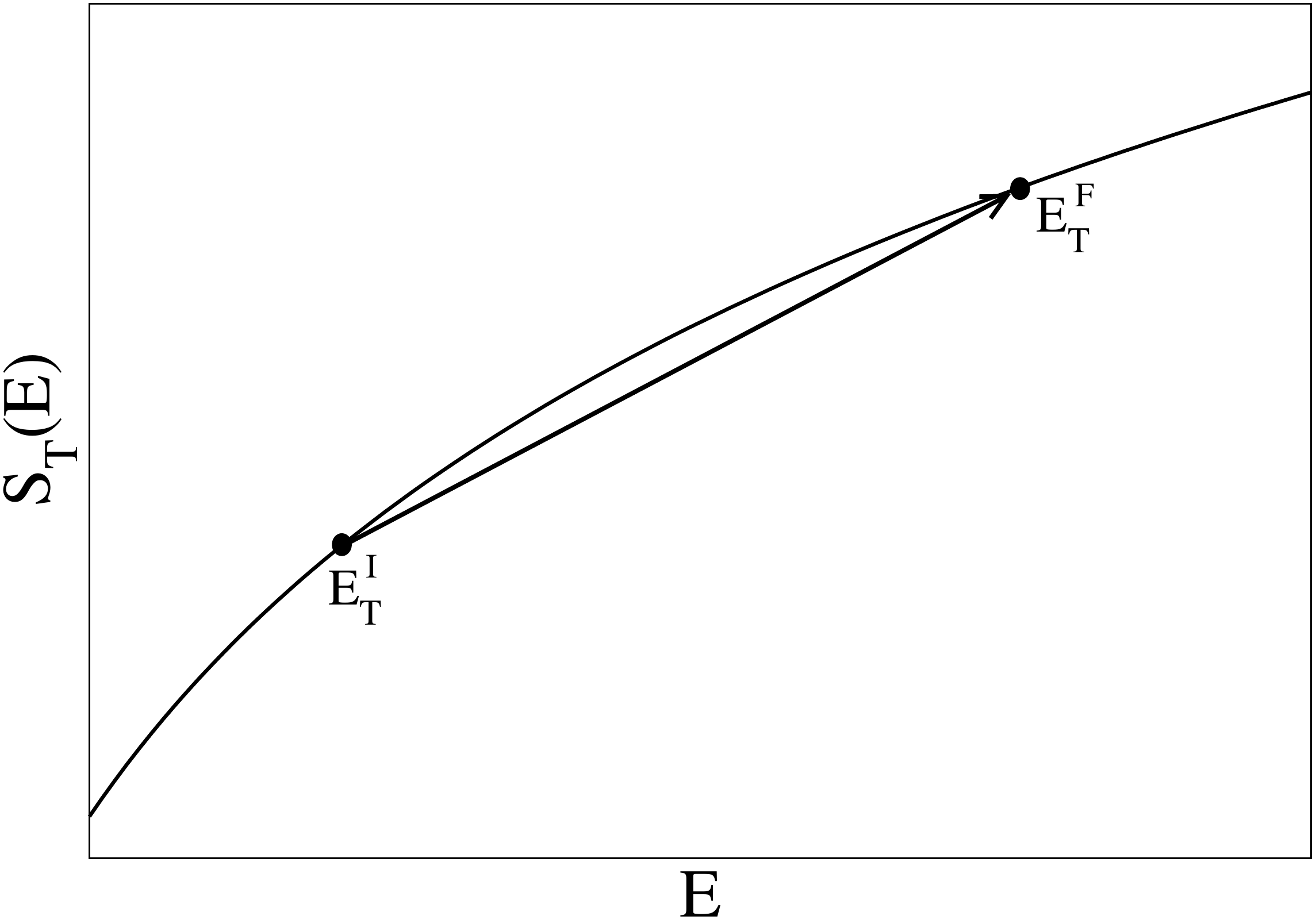}
\caption{Entropy of the system $S_S(E)$ (left panel), and of the
  thermometer $S_T(E)$ (right panel), as a function of $E$. At initial
  time the system's energy $E_S^I$ corresponds to a negative
  temperature, $T_I<0$; after the thermalization, due to the coupling
  with the ``standard'' thermometer, the system's temperature must be
  necessarily positive, and a huge transfer of energy from the system
  to the thermometer occurs, in such a way that $E_S^I+E_T^I \simeq
  E_S^F+E_T^F$.}
\label{fig:entropy}
\end{figure}

It is useful to recall an argument to show that the coupling of the system
$\mathcal{A}$ at negative temperature with a system $\mathcal{B}$
which can have only positive temperature always produces a system with
final positive temperature.  Indeed, at the initial time the total
entropy is
\begin{equation}
S_{I}=S^\mathcal{A}(E_\mathcal{A})+S^\mathcal{B}(E_\mathcal{B}),
\end{equation}
while, after the coupling, it will be 
\begin{equation}
S_{F}=S^\mathcal{A}(E'_\mathcal{A})+S^\mathcal{B}(E'_\mathcal{B}),
\end{equation}
where $E'_\mathcal{A}+E'_\mathcal{B}=E_\mathcal{A}+E_\mathcal{B}$ and, within our assumptions, $E'_\mathcal{A}$ is
determined by the equilibrium condition that $S_{F}$ takes the maximum possible
value~\cite{huang}, i.e.
\begin{equation}
\beta_\mathcal{A}={\partial S^\mathcal{A}(E'_\mathcal{A}) \over \partial E'_\mathcal{A}}=
\beta_\mathcal{B}={\partial S^\mathcal{B}(E'_\mathcal{B}) \over \partial E'_\mathcal{B}}.
\end{equation}
Since $\beta_\mathcal{B}$ is positive for every value of
$E_\mathcal{B}'$, the final common temperature must also be
positive. The above result helps to understand why it is not common to
observe negative temperatures.  Therefore using as thermometer a
``standard'' system with only positive $\beta$ as those used for the
FPU system, the thermometer cannot work.  In Fig.~\ref{fig:entropy} we
show schematically the mechanism for the energy transfer.

\begin{figure}[ht!]
\centering
\includegraphics[width=0.49\linewidth]{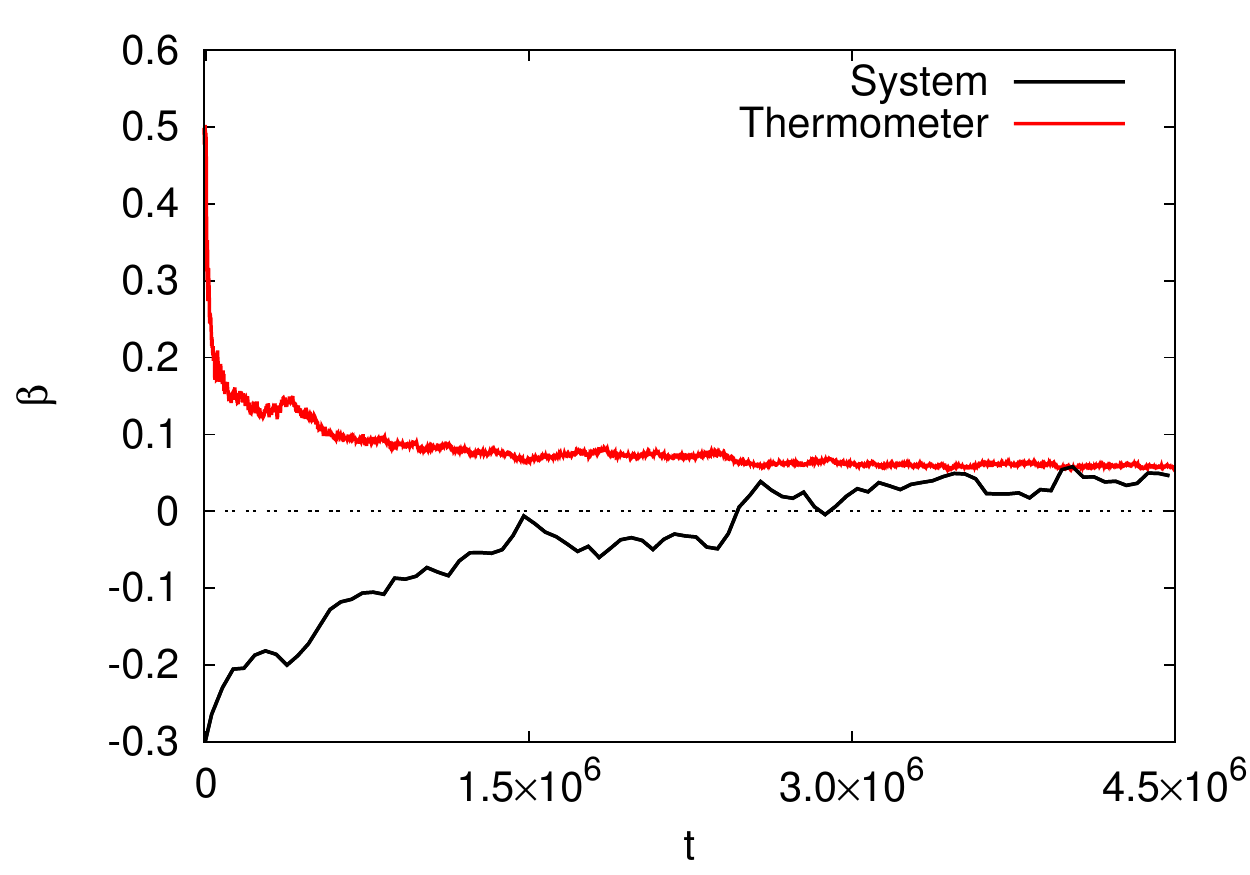}
\includegraphics[width=0.49\linewidth]{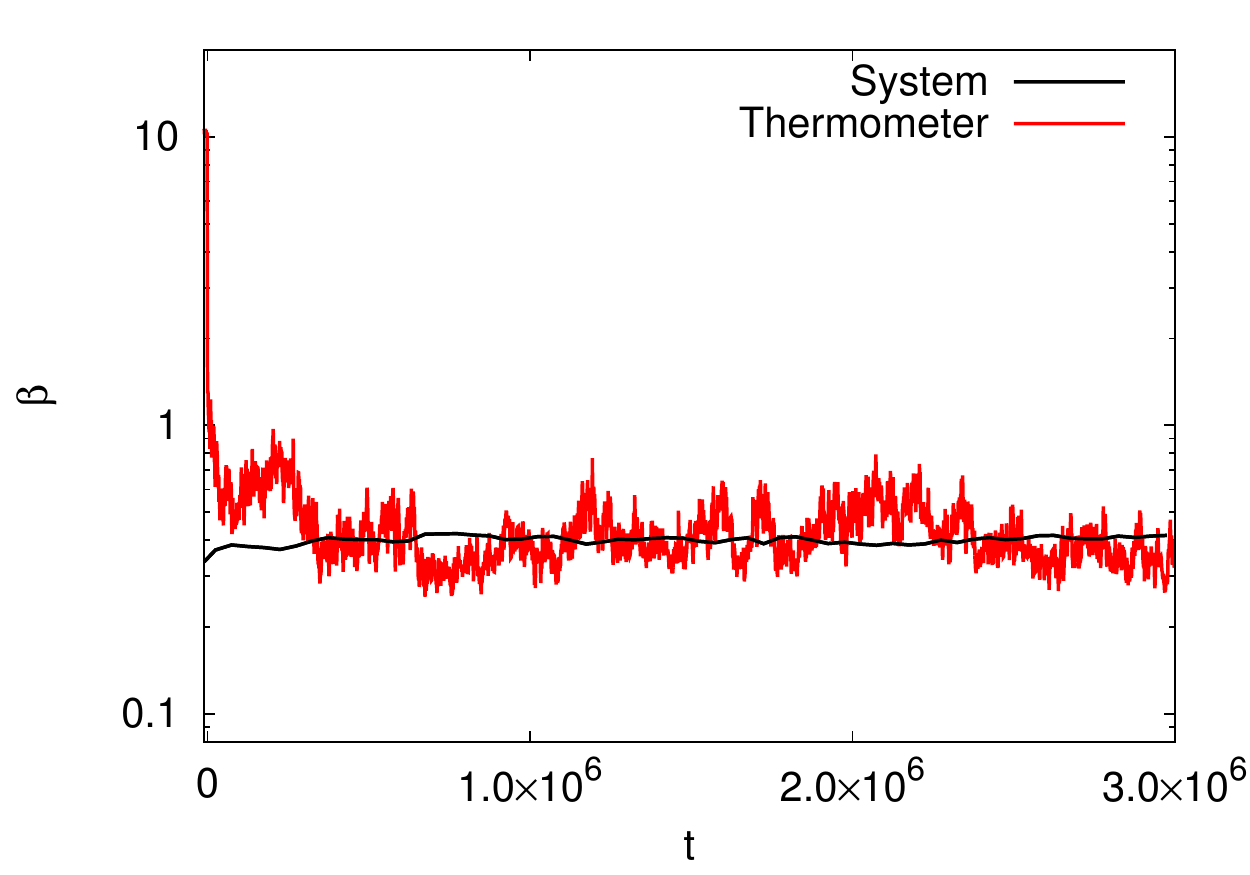}
\caption{Inverse temperature $\beta$ as a function of time, for the
  system \eqref{eq:tn} and the ``usual'' thermometer
  \eqref{eq:term_ang} (with $m=1$). In the left (right) panel the
  $\beta$ corresponding to the initial energy is negative (positive).
  $\beta(t)$ of the system is computed from a fit on the single
  particle momentum p.d.f.: we consider the histogram of the measured
  momenta from time $t$ to time $t+\Delta t$, with $\Delta t=50000$ in
  this case, and we get the value of $\beta$ from the slope of
  $\log[\rho(p)]$, as explained in Ref. \cite{cerino15}. Thermometer's
  inverse temperature has been determined, as usual, by $\beta=\langle
  p^2/m\rangle ^{-1}$. Parameters: $N_S=1000$, $N_T=30$,
  $K=\gamma=0.5$, $J=0.05$ $\epsilon=0.1$.}
\label{fig:tp_time}
\end{figure}

\begin{figure}[ht!]
\centering
\includegraphics[width=0.49\linewidth]{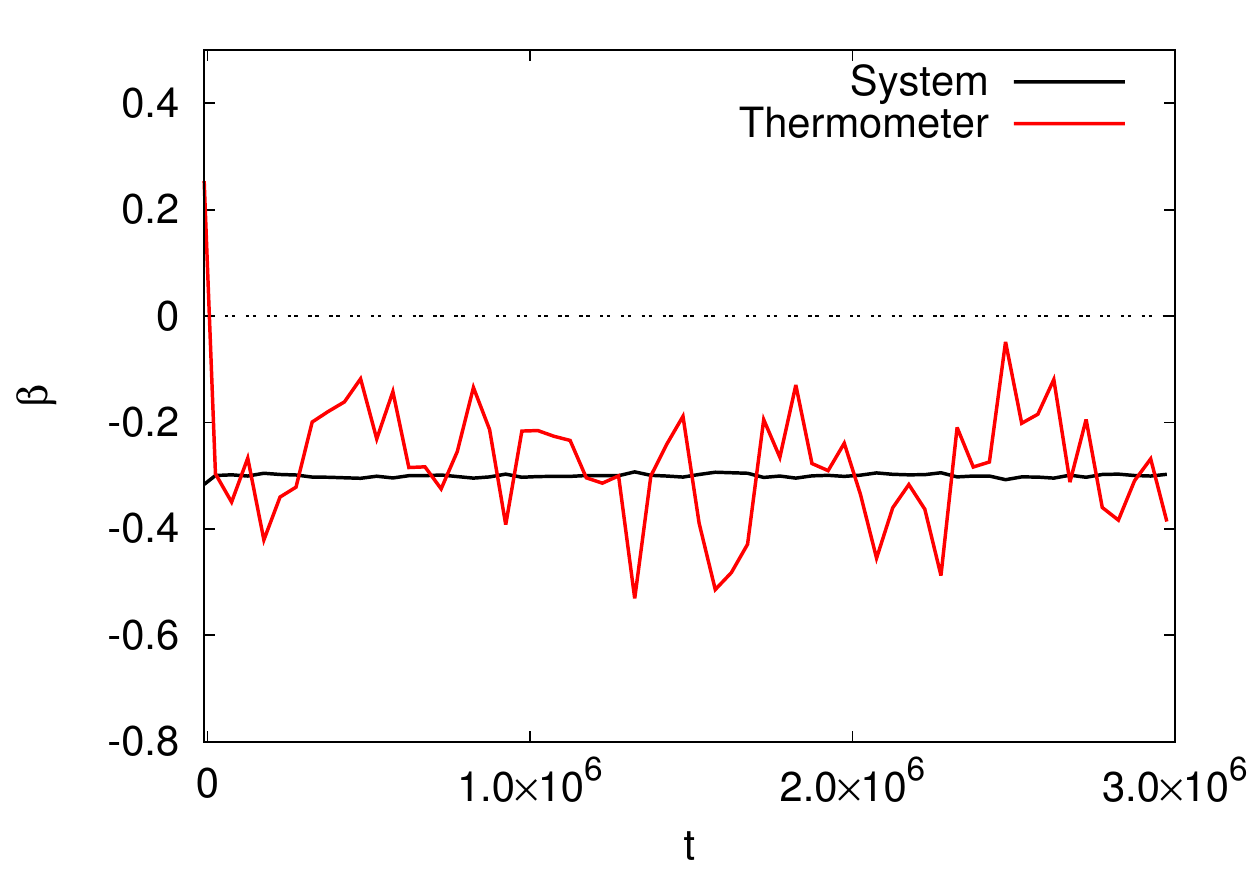}
\includegraphics[width=0.49\linewidth]{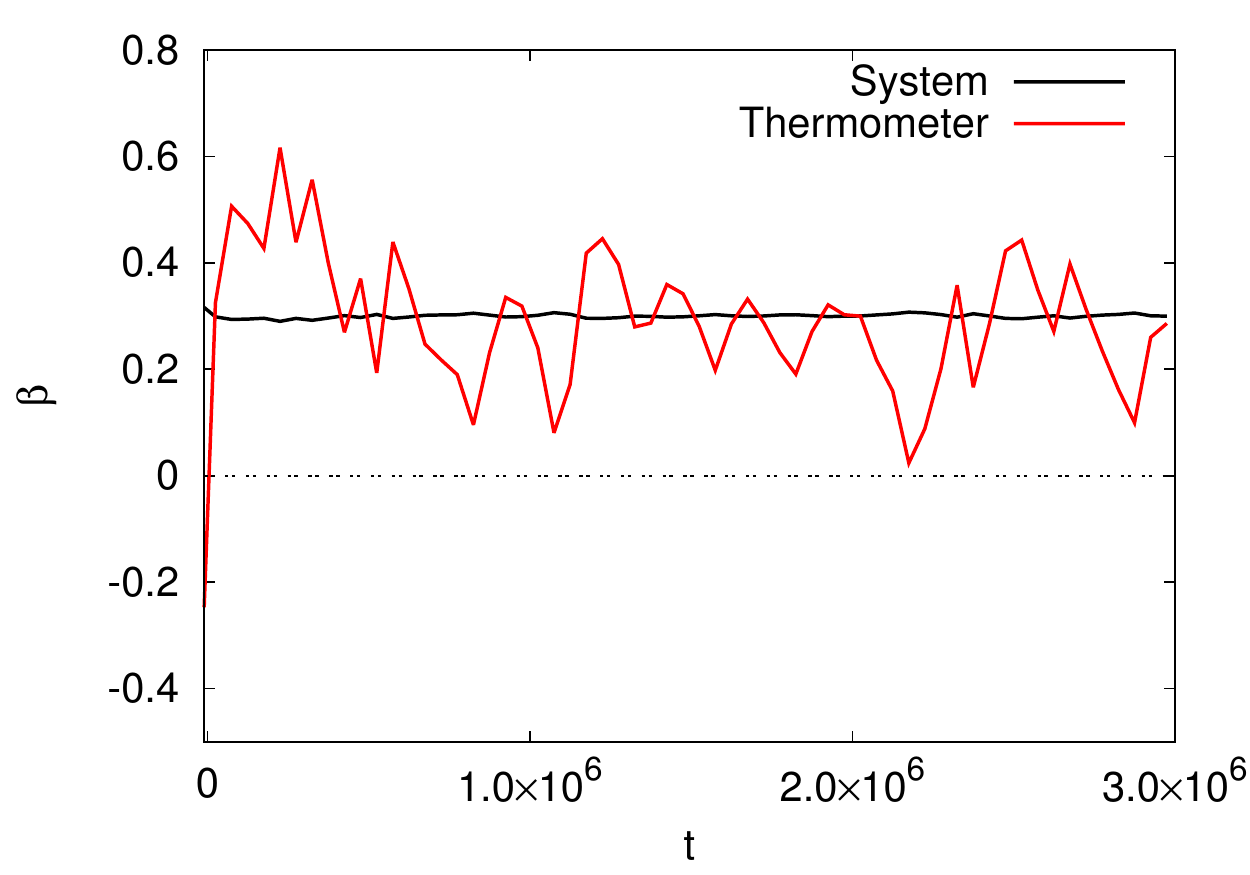}
\caption{Inverse temperature $\beta$ as a function of time, for the
  system \eqref{eq:tn} and the thermometer \eqref{eq:tn_term}, both 
  measured by the distribution fitting procedure explained in Fig~\ref{fig:tp_time}. In the left (right) panel the
  $\beta$ corresponding to the initial energy is negative (positive). Parameters as in Fig. \ref{fig:tp_time}. }
\label{fig:tn_time}
\end{figure}

We have checked this argument using the ``standard'' thermometer with Hamiltonian given by Eq.~\eqref{eq:term_ang} and the following interaction Hamiltonian
\begin{equation}
 H_I=\sum_{i=1}^{N_T} \left[1-\cos(\phi_i-\theta_i)\right].
\end{equation} 
Figure~\ref{fig:tp_time} shows the scenario predicted by the previous
simple thermodynamic arguments: in the left plot we see how starting from a situation at negative
temperature we have an energy flux from the system to the thermometer.
The amount of the exchanged energy is huge even if the size of the
thermometer is small (and its coupling $\epsilon \ll 1$) and the final
state of the system must be very different from the initial one.  Therefore
we can say that the thermometer acts as a ``vampire'' producing a
qualitative change in the system.

On the contrary using a proper thermometer able to measure even
negative temperature, i.e. one with Hamiltonian 
\begin{equation}
\label{eq:tn_term}
H_T=\sum_{i=0}^{N_T}(1-\cos p_i)+\sum_{i=1}^{N_T}\gamma\left[1-\cos(\phi_i-\phi_{i-1})\right] 
\end{equation}
a correct measurement is achieved, see Fig.~\ref{fig:tn_time}.

\section{Conclusions}

We have discussed the meaning of temperature and the issue of defining
a proper model thermometer in non standard cases, featuring systems
with long-range interactions or bounded phase space.  Starting from
the generalized Maxwell-Boltzmann distribution, allowing one to
measure the temperature as a time average for Hamiltonian systems with
a generic form of the kinetic part, we have considered different
mechanical models for a thermometer. First, we have studied the case
of FPU-like chains, showing that, even in the presence of weak
ergodicty at low energy, at equilibrium a thermometer coupled to the
system measures the proper temperature. Second, we have considered the
generalized Hamiltonian mean field model, characterized by long-range
interactions. In this case, we have introduced a model thermometer
with angular variables and we have shown that it efficiently
determines the system temperature. Finally, we have addressed the
interesting issue of the measurement of temperature in systems with a
bounded phase space, where such a quantity can take on negative
values.  For these systems we have shown that, in order to measure in
a correct way the system's temperature, one has to introduce a
suitable model thermometer, with a negative temperature scale.

\section*{Appendix: Equilibrium properties of a ``hybrid'' system for $\beta<0$}

Let us consider the system described by the Hamiltonian 
\begin{eqnarray}
 H&=&\sum_{i=1}^N{(1-\cos p_i) +N\frac{J}{2}\left[ 1-\left(\frac{1}{N}\sum_{i=1}^N \cos \theta_i \right)^2 - \left(\frac{1}{N} \sum_{i=1}^N \sin \theta_i\right)^2 \right]} \nonumber \\
&+& K \sum_{i=1}^{N} [1-\cos(\theta_{i+1}-\theta_i)]\label{eq:ham}
\end{eqnarray}
with $(\theta_{N+1} \equiv \theta_1)$. In order to determine its
equilibrium properties we need to find the canonical equipartition
function $Z(\beta,N)$, and then to compute the free energy per
particle through the limit $-\beta f(\beta)=\lim_{N \rightarrow
  \infty} \log Z/N$.

$Z(\beta, N)$ can be factorized into two terms:
\begin{itemize}
 \item a kinetic part,
 \begin{equation}
 \label{eq:kin_app}
   Z_K(\beta, N)=\int_{-\pi}^{\pi}dp_1...dp_N \exp \left[ -\beta \sum_{i=1}^N(1-\cos p_i) \right];
 \end{equation}
 \item a configurational one, 
 \begin{equation}
 \label{eq:conf_app}
   \begin{aligned}
   Z_C(\beta, N)=&e^{-\beta J N/2} \int_{-\pi}^{\pi} d\theta_1 ... d\theta_N \exp \left[\frac{\beta J}{2N}\left(\sum_{i=1}^N \cos \theta_i\right)^2 + \frac{\beta J}{2N}\left( \sum_{i=1}^N\sin \theta_i\right)^2\right]\\ \times&\exp\left[ -\beta K \sum_{i=0}^N (1-\cos(\theta_{i+1}-\theta_i))\right].
 \end{aligned}
 \end{equation}
\end{itemize}
The kinetic contribution to $f(\beta)$ can be easily computed from \eqref{eq:kin_app} in terms of the Modified Bessel Functions of the first kind, as in Ref.~\cite{cerino15}; one gets
$$
-\beta f_K(\beta)\equiv\lim_{N \rightarrow \infty}\frac{1}{N} \log Z_K = -\beta + \log \left(2 \pi I_0(\beta) \right).
$$
The strategy to determine the configurational part has been outlined in Ref.~\cite{ruffo1995}, where a model with a similar interacting potential has been extensively studied. The result for $\beta >0$ is
\begin{equation}
\label{eq:f_bpos}
-\beta f_C(\beta)\equiv\lim_{N \rightarrow \infty}\frac{1}{N} \log Z_C = -\beta K - \beta \frac{J}{2} - \min_{m\ge0} \left(\frac{m^2}{2\beta J} -\log[\lambda(m, K \beta)]\right)
\end{equation}
where $\lambda(z,\alpha)$ is the maximum eigenvalue of the symmetric integral operator
\begin{equation}
 (\mathcal{T}_{z,\alpha} \psi)(\theta)=\int_{-\pi}^{\pi} d \theta' \exp \left[\frac{1}{2}z(\cos \theta + \cos \theta ') + \alpha \cos(\theta-\theta') \right]\psi(\theta').
\end{equation}
In the following we will explicitly derive the case $\beta<0$, with the same strategy which has been used in Ref.~\cite{ruffo1995} for $\beta>0$.

\begin{figure}[ht!]
\centering
\includegraphics[width=0.5\linewidth]{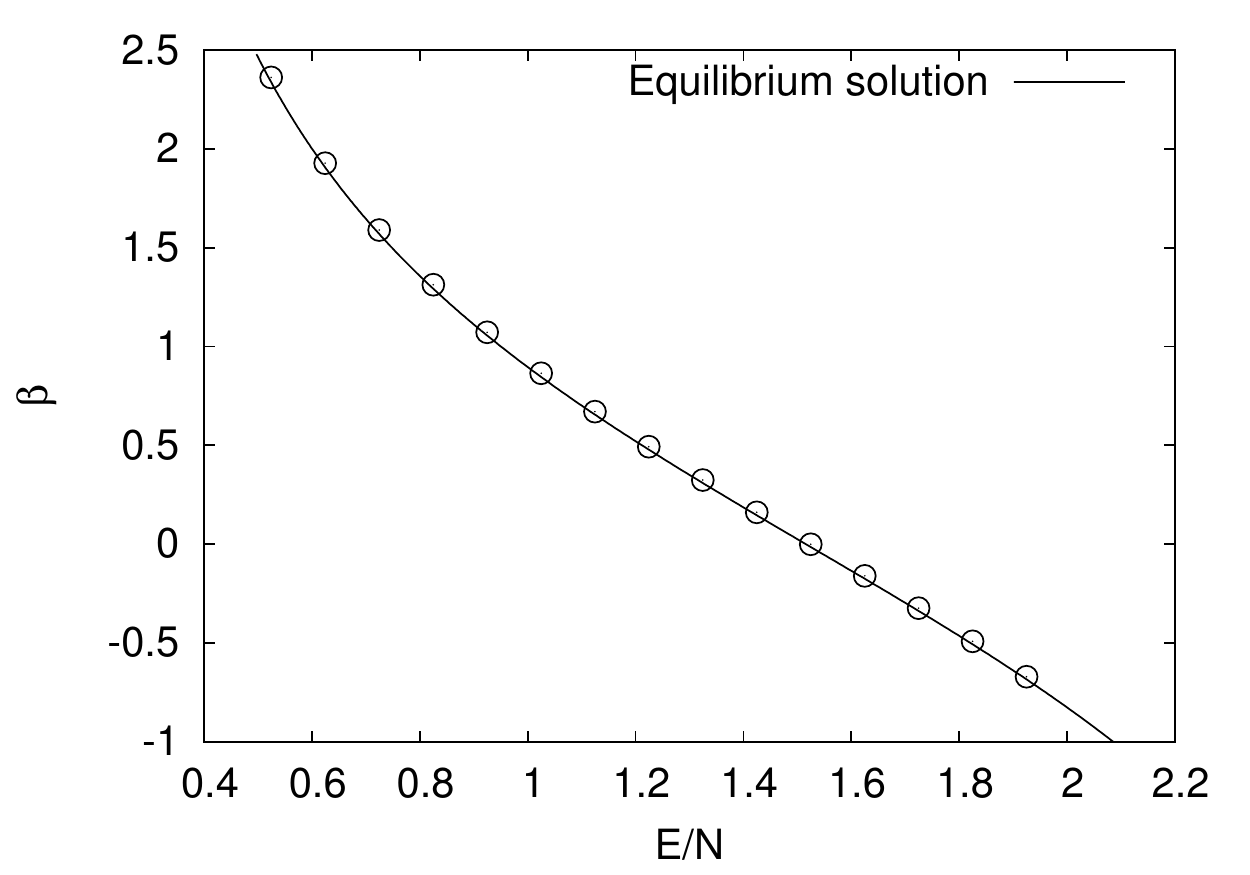}
\caption{Equilibrium behaviour of the system~\eqref{eq:tn}. Analytical
  solution (solid line) and simulations with $N=200$ (circles) are
  compared, for $K=0.5$, $J=0.05$.}
\label{fig:tn}
\end{figure}

From Eq. \eqref{eq:conf_app}, by mean of a standard Hubbard-Stratonovich transformation, we get
 \begin{equation}
 \label{eq:hub-strat}
   \begin{aligned}
   Z_C(\beta, N)&=\frac{N e^{-\beta J N/2}}{2 \pi|\beta| J}\int d \rho_x d\rho_y \exp\left[-\frac{N}{2|\beta|J}(\rho_x^2+\rho_y^2)-N \beta K \right] \times\\
   & \times\int_{-\pi}^{\pi} d\theta_1...d\theta_N \exp \left[i \sum_{j=1}^N (\rho_x \cos \theta_j + \rho_y \sin \theta_j) + \beta K \sum_{j=1}^N \cos (\theta_{j+1}-\theta_j)  \right]=\\
   & =\frac{N e^{-\beta J N/2}}{2 \pi|\beta| J} \int_0^{\infty} d \rho \int_{-\pi}^{\pi}d\phi\, \rho \exp\left[-N\frac{\rho^2}{2|\beta|J}-N\beta K \right] \times\\
   & \times\int_{-\pi}^{\pi} d\theta_1...d\theta_N \exp \left[ i \rho \sum_{j=1}^N  \cos (\theta_j - \phi) + \beta K \sum_{j=1}^N \cos (\theta_{j+1}-\theta_j)  \right],
 \end{aligned}
 \end{equation}
 where we have introduced ``polar'' coordinates for the plane $(\rho_x, \rho_y)$.
Now we shift all the integration variables $\{\theta_j\}$ by an angle $\phi$, then we can rewrite the last term in a symmetric fashion and recover the functional form of the integral operator $\mathcal{T}$ introduced above, so that
 \begin{equation}
 \label{eq:trace}
 \begin{aligned}
   Z_C(\beta, N)&=\frac{N e^{-\beta J N/2}}{2|\beta| J} \int_0^{\infty} d \rho^2 \, \exp\left[-N\frac{\rho^2}{2|\beta|J}-N\beta K\right] Tr\left[\mathcal{T}_{i\rho,\beta K}^N   \right]\\
   &\simeq\frac{N e^{-\beta J N/2}}{2|\beta| J} \int_0^{\infty} d \rho^2 \, \exp\left[ -N\frac{\rho^2}{2|\beta|J} - N\beta K+N\log[\lambda(i\rho, K \beta)] \right]
 \end{aligned}
 \end{equation}
 (reminding $\theta_1 \equiv \theta_{N+1}$). The last equality holds in thermodynamical limit, $N\gg 1$. Note that since $\mathcal{T}_{i\rho,\beta K}$ is not an hermitian operator, its eigenvalues will be in general complex numbers: therefore $\lambda(z,\alpha)$ has to be defined, in this case, as the eigenvalue with the maximum modulus. Finally, one can use steepest-descent method to reduce the calculation of the integral to a minimization problem, that can be solved numerically. 
\\
In Fig.~\eqref{fig:tn} we compare our analytical prediction for the curve $\beta(E)$ with computer simulations of the system at equilibrium, finding an excellent agreement.

\section*{Bibliography}

\bibliographystyle{iopart-num}
\bibliography{biblio}

\end{document}